\documentclass[preprint,12pt]{elsarticle}
\biboptions{authoryear,sort}

\usepackage{epsfig}
\usepackage{epstopdf}
\usepackage{ulem}
\usepackage{color}

\usepackage{amssymb}

\newcommand{\Facc}{Fermi acceleration}
\newcommand{\syn}{synchrotron}

\newcommand{\Kep}{K_\mathrm{ep}}
\newcommand{\gamray}{$\gamma$-ray}
\newcommand{\gamrays}{$\gamma$-rays}
\newcommand{\piZ}{\pi^0}
\newcommand{\pcc}{cm$^{-3}$}

\newcommand{\csps}{cm$^2$ s$^{-1}$}
\newcommand{\Ice}{{\sl IceCube}}
\newcommand{\Hess}{H.E.S.S.}
\newcommand{\CXO}{CXO J164710.2-455216}
\newcommand{\WestI}{Westerlund~1}
\newcommand{\WestII}{Westerlund~2}
\newcommand{\Emax}{E_\mathrm{max}}
\newcommand{\kmps}{km s$^{-1}$}
\newcommand{\Tacc}{T_\mathrm{acc}}
\newcommand{\lsim}{\lesssim\!}
\newcommand{\gsim}{\gtrsim\!}
\def\apj{ApJ}
\def\apjl{ApJ Lett}
\def\mnras{MNRAS}
\def\nat{Nature}

\def\prl{PRL}
\def\PRL{PRL}
\def\araa{ARA\&A}                
\def\aap{A\&A}                   

\def\prd{Phys. Rev. D}

\def\ssr{Space Sci. Rev.}
\def\aapr{Astron. Astroph. Reviews}

\def\jcap{JCAP}
\def\APP{Astro. Part. Phys.}

\newcommand{\xx}[1]{\!\times\!10^{#1}}

\usepackage[ps2pdf,%
a4paper=true,%
breaklinks=true,%
colorlinks=true,%
pdfauthor={First Author et al.},%
pdftitle={Template for manuscripts in Advances in Space Research}%
]{hyperref}

\journal{Advances in Space Research}

\begin{document}

\begin{frontmatter}



\title{Supernovae in compact star clusters as sources of high-energy cosmic rays and neutrinos}




\author{Bykov A.M.$^{1,2}$}
\ead{byk@astro.ioffe.ru}
\author{Ellison D.C.$^{3}$}
\ead{ellison@ncsu.edu}   
\author{Gladilin P.E.$^{1}$}
\ead{peter.gladilin@gmail.com}
\author{Osipov S.M.$^{1}$}
\ead{osm2004@mail.ru}
\address{$^{1}$Ioffe Institute, Saint-Petersburg, Polytechnicheskaya str., 26, 194021, Russia}
\address{$^{2}$ International Space Science
  Institute, Bern, Switzerland}
\address{$^{3}$North Carolina State University, Department of Physics, Raleigh, NC 27695-8202, USA}

\begin{abstract}
We discuss a specific population of galactic PeVatrons which may be
the main source of the galactic (cosmic-ray) (CR) component well above PeV energies.
Supernovae in compact clusters of massive stars are proposed as
powerful sources of CRs, neutrinos, and
\gamray\ emission.
Numerical simulations of non-linear \Facc\ at converging
shock flows have revealed that these accelerators can provide very
hard spectra of protons up to $10^{16}-10^{17}$ eV which is well
above the ``knee'' in the all-particle CR spectrum at about
$3\times10^{15}$ eV. We suggest that known supernova remnants
interacting with stellar winds in the compact clusters of young
massive stars Westerlund I and Cl*1806-20 can be associated with the
sources of the TeV \gamray\ emission detected by H.E.S.S. and may be
responsible for a fraction of the high-energy neutrinos detected
with the {\sl IceCube} observatory. A recent CR
composition measurement with the {\sl LOFAR} array has revealed a
light-mass component possibly dominating the all-particle spectrum
at energies around $10^{17}$ eV. Such a strong light component
(mainly protons and helium) may require  specific galactic CR
sources  such as supernovae interacting with compact clusters of
massive stars in addition to isolated supernova remnants.
\end{abstract}


\end{frontmatter}
\section{Introduction}
Measurements of cosmic ray (CR) composition at energies
just above the spectral knee \citep[see e.g.,][]{KASCADE13,
berezhnevea13,Buitink2016} are of fundamental importance for deducing the origin of CRs
\citep[e.g.,][]{hillas05,Strong2007,blasiAARv13,amato14,blandford14}.
Recent analysis of the atmospheric depth of CR shower maxima
measured with the Low Frequency Array (LOFAR) \citep{Buitink2016}
have provided evidence for a predominantly light-mass CR composition
at energies around $10^{17}$ eV, i.e., well above the spectral steepening
at $\sim 3\xx{15}$\,eV referred to as the ``knee."

The knee in the all-particle CR spectrum is believed to result from a rigidity (i.e., charge) dependent cut-off in shock acceleration occurring in isolated
supernova remnants (SNRs). If this is the case, heavy elements obtain a higher energy with the iron cut-off reaching 26 times that of protons. The overall composition in the knee region should become heavier with energy and be dominated by heavy elements around $10^{17}$\,eV.
If the LOFAR observations showing a large light-mass fraction are confirmed, an additional source of light nuclei may be needed as we consider here.

To date several models of particle accelerators which are able to
accelerate CRs beyond the knee have been suggested. Some of them
consider CR acceleration at the galactic wind termination shock
\citep[e.g.,][]{1987ApJ...312..170J}, while others deal with
individual isolated supernovae with varying explosion energies
\citep[e.g.,][]{2003A&A...409..799S} and in
different circumstellar environments
including exploding Wolf-Rayet (WR) stars in
stellar cavities \citep{1988ApJ...333L..65V}, core-collapse
supernovae (SNe) with dense circumstellar winds \citep{Ptu10}, or type
Ibc supernovae with trans-relativistic shocks
\citep[e.g.,][]{MCtransrel13}.

Models assuming the collective effects of multiple clustered
SNe and strong winds of young massive stars in galactic
superbubbles produced by OB-star associations have also been
considered by \citet{bt01,b01,parizot04,fermarcow10,b14}. These models
all allow the acceleration (or re-acceleration) of CRs to beyond the
knee. In any case, recent \gamray\ observations of the Cygnus
Superbubble \citep{fermiSB11} and a candidate young massive OB
association/cluster G25.18+0.26 \citep{G25} in the
Milky Way, as well
as in starburst galaxies \citep[see
e.g.,][]{b14,2016CRPhy..17..585O}, have provided confirming evidence
that SNe can be efficient CR accelerators in such an
environment.

It's clear that many SNe occur in compact  star clusters.
Such clusters or superclusters of mass $>10^5\,M_{\bigodot}$ may
contain hundreds or thousands of young massive stars within a parsec
size core \citep[see
e.g.,][]{Clark2008,massive_star_cluster_2010ARA&A}.
Such systems are often
observed in external starburst or interacting galaxies while a few
of them have been found in Milky Way-type galaxies. The most massive
stars in such a cluster must begin to explode just a few million years
after the cluster is born.
Interacting with multiple stellar winds in a cluster environment,
core-collapse SNe can accelerate particles to well above
PeV energies in less than a thousand years  \citep{b14,Bykov2015}.
A specific feature of such systems, predicted by non-linear
modeling, is that the particles accelerated therein have very hard
spectra at PeV energies.

\section{Supernovae with colliding shock flows as efficient galactic PeVatrons}
The most massive stars in compact clusters explode as core collapse
supernovae after a few million years. Their shocks will travel
through regions filled with fast winds from other massive young
stars, or with the global cluster wind, forming
{\it colliding shock flow} (CSF)
systems. These CSF systems can result in an efficient conversion of
the kinetic power of a supernova shell and stellar winds into the
power of high-energy CRs.
The first-order Fermi mechanism accounting for the non-linear
backreaction of the accelerated CRs on the colliding shock structure
was discussed in detail in \cite{MNRAS_BGO13,Bykov2015}. In these
parsec-sized systems, the CSF stage starts a few hundred years
before the SN shock collides with the wind termination shock. At
this time, the maximum energy particles that have been accelerated
at the SNR shock are able to reach the fast wind termination shock
and can be scattered back to the SNR by magnetic fluctuations
carried by the fast stellar wind. Therefore, high-energy particles
with mean free paths larger than the distance between the two shocks
start to be accelerated by the converging fast flows. This is the
most favorable circumstance for efficient Fermi acceleration. It
results in a high pressure of accelerated particles at the highest
energy end of the CR distribution, which decelerates the colliding
flows.

Simulations predict non-power-law spectra with a strong upturn above several TeV becoming extremely hard at PeV energies.
The upturn occurs when particles are accelerated to an energy allowing them to scatter between the SNR shock and the stellar wind shock thereby gaining energy in a fashion more efficient than in a single shock.

\begin{figure}    
\resizebox{12cm}{10.cm}{
\includegraphics[trim= 100 240 160 310,clip]{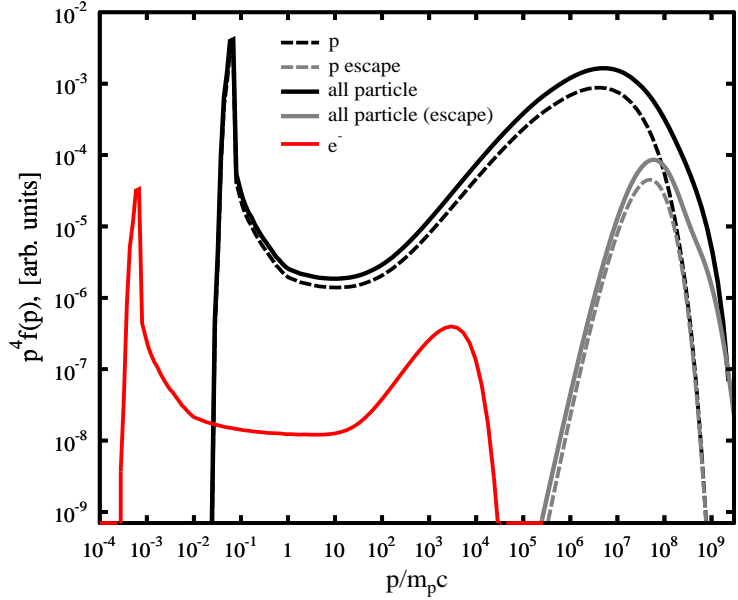}}
\caption{Cosmic rays accelerated by a colliding shock flow (CSF) system at the end of the CSF phase. The all-particle hadron spectrum (i.e., the sum of protons and heavier nuclei) is shown
by a thick black line for the trapped CRs and by a gray line for the escaping CRs.
The electron spectrum (lowest curve), which is suppressed by synchrotron losses at $\gsim 5$\,TeV, is shown in red.
The troughs in the spectra below $\sim 100$\,GeV result when acceleration is dominated by the isolated SNR shock.  The harder spectral shapes at higher energies are due to the efficient particle acceleration when the SNR CRs interact with the colliding stellar wind shock
\citep[see][]{MNRAS_BGO13, gladilin14}.
The primary electron-to-proton ratio at relativistic energies is set by the normalization factor $\Kep=0.03$ \citep[e.g.,][]{Park2015}.}
\label{particles}
\end{figure}

It was shown that during the period of the most efficient
acceleration (up until several hundred years after the SN explosion)
the intensity of CR production in a CSF system
exceeds that of an isolated SNR in the 10$^{14}$ -- 10$^{17}$ eV regime. Due to the
fast magnetic field amplification by CR driven instabilities and
efficient CR acceleration in the colliding shock flows, the spectrum
of accelerated particles has a much shorter evolution time than that
in an isolated SNR \citep{Bykov2015}.
In Fig.~\ref{particles} both proton and electron spectra are shown
at the end of the CSF acceleration phase, just before the shocks
collide. The hard spectrum expected from CSFs puts most of the
energy into the highest energy protons. This feature of the spectral
energy distribution leads to high fluxes of \gamrays\ in the range
between 10$^{14}$--10$^{16}$ eV [see red (lowest) curve in
Fig.~\ref{particles}]. Also, CSF systems, along with the remnants
of IIb and IIn  \citep{Ptu10} or Ibc SNe
\citep{MCtransrel13}, are unique galactic accelerators which are
able to contribute importantly to the overall CR spectrum at
energies above the ``knee'' ($3\cdot10^{15}$ eV).

The maximal energy of a particle accelerated at a CSF system depends
on the velocities of the colliding plasma flows, on details of the
system geometry,  and on the magnitude of the magnetic field. The
proton acceleration time in a CSF system was estimated in
\citet{Bykov2015} as $\Tacc \approx 2\,\cdot10^{10}\, \epsilon_{\rm
PeV}\, (\eta_b n)^{-0.5}\,u_{s3}^{-2}\,u_{w3}^{-1}$ s. Where $\eta_b
<$1 is the magnetic field amplification efficiency, the CR proton
energy $\epsilon_{\rm PeV}$ is in PeV, and the speeds of supernova
ejecta $u_{s3}$ and cluster wind $u_{w3}$ are in units of
$10^3$\,\kmps. For typical parameters of a young SN in a cluster
this results in a maximal energy of accelerated CRs $\Emax \sim
Z\cdot2\cdot 10^{16} \,\mathrm{eV}$, where $Z$ is the charge of the
particle (see Fig.~\ref{particles}). 
The value of $\Emax$ can
be much higher than the above estimate if the CSF system was initiated by
a hypernova in a cluster of young massive stars. In any case, $\Emax$
exceeds the maximal CR energy for typical isolated SNRs.

\section{The contribution of supernovae in stellar clusters to the CR spectrum}

The unique features of CSF systems of having particularly hard spectra and short acceleration lifetimes ($\tau_a\leq500$ yrs) make it possible for these sources to
dominate at PeV energies without violating observational constraints
at lower CR energies.

\begin{figure}            
\begin{center}
\resizebox{15cm}{9.5cm}{
\includegraphics[trim=140 290 100 315 ,clip]{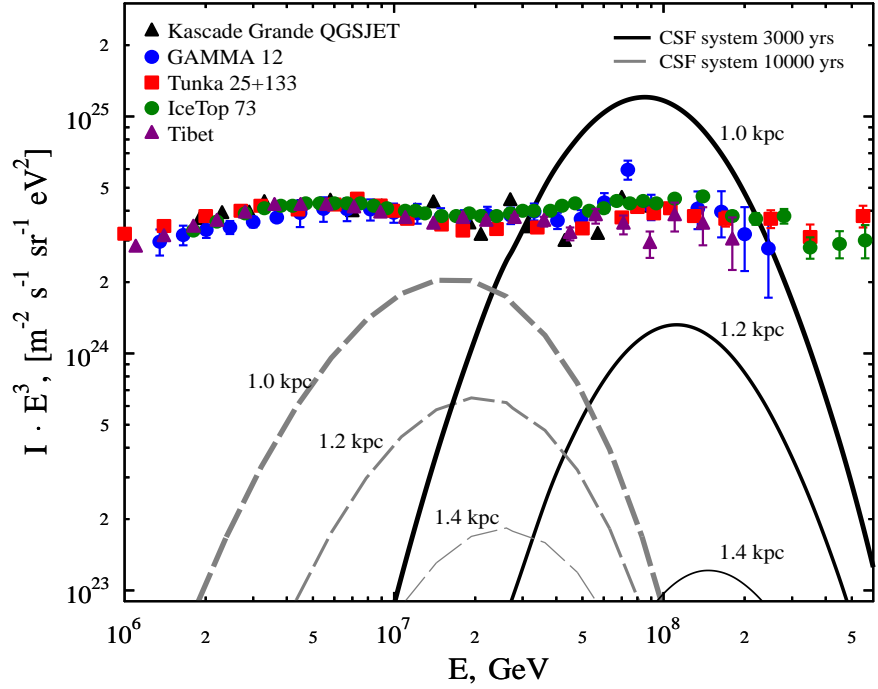}}
\caption{All-particle energy spectrum: data points from Tunka-25,
KASCADE-Grande, Tibet-III, GAMMA, Tunka-133 and IceTop-73
\citep{Prosin2014, aartsen2013, Garyaka13,
apel2009,amenomori2008,Budnev2013}
along with the calculated all-particle CR
flux from a 3000 year old CSF source \textit{(black lines)} and from a
$10^4$ year old CSF source \textit{(dashed gray lines)} for the distances from the Earth as indicated.
}
\label{DataSpectrum}
\end{center}
\end{figure}

\subsection{Single CR source contribution}
Consider a system where a single supernova shell is colliding with the
stellar wind of a nearby massive star. For a few hundred years the
system will be an efficient accelerator: it will produce high-energy
particles bouncing between the converging flows.
Fig.\ref{particles} shows trapped CR spectra and escaping fluxes for such a system where non-linear effects and magnetic field
amplification have been modeled.

In the terms of the halo-diffusion model, the flux of nuclei with
charge $Z$ coming from a single source located at the distance $d$
can be evaluated as follows \citep[see e.g.][]{Sveshnikova2013}:
\begin{eqnarray}
F(E) =
\frac{c}{4 \pi}
S(E,Z)
\frac{\exp{\left(-\frac{d^2}{4 R_d^2}\right)}}{4 \pi H R_d^2 }\times \nonumber \\
\times\sum_{n=0}^{\infty}
\exp{\left[-\frac{(2n-1)^2 \pi^2 R_d^2 }{4 H^2}\right]},
\end{eqnarray}
%
%
where $S(E,Z)$ is the flux escaping from the source
(see Fig.~\ref{particles}), $R_d= \sqrt{D(E/Z)t}$ is
the diffusion radius of a particle, $H=$4~kpc is the width of the Galactic halo,
$t$ is the age of the source, and
$D=5\cdot10^{29}(E/\mathrm{PeV})^{0.33}$\,\csps\ is the diffusion coefficient.

The total energy of a CSF source contained in CRs can be as high as
2$\cdot$10$^{49}$ erg. A typical lifetime of a CSF system is $\sim
300$ years, so the $\gamma$-ray emission related to the SNR and its
neighborhood is time-variable on scales of tens of years.
Thus, if CSF systems actually contribute to the observed CR flux at
energies beyond the knee,  the \gamray\ emission from these systems
will be time-variable over observable epochs.

In Fig.~\ref{DataSpectrum} we show  simulated all-particle  spectra
from CSF sources at 1.0, 1.2, and 1.4 kpc distances for young (3,000
yr) and relatively old (10,000 yr) systems.
The modeling assumes the sources accelerate particles up to
$Z\cdot$ 40~PeV.
Since the acceleration mechanism depends on the CR rigidity, $R=pc/Ze$,
the maximal energy of the nuclei is proportional to charge.
The velocity of the SNR shock is assumed to be 6000\,\kmps, the
velocity of the stellar wind is assumed to be 4000\,\kmps, and the
magnetic field is assumed to be amplified at the shock up to
900~$\mu$G.

Fig.~\ref{DataSpectrum} shows that a single, relatively young CSF
source can make a significant contribution to the observed CR
spectrum if it is within $\sim 1.2$\,kpc. A young SNR (age $\lsim
10$\,kyr) is required both to accelerate nuclei up to the highest
energies and to provide the highest possible CR fluxes. Our
simulations have shown that a source  within 1.4\,kpc could
contribute $5-10\%$ of the total CR flux between $\sim 3\xx{16}$ and
$10^{17}$\,eV, while one within 1.2\,kpc could contribute $\gsim
30\%$ and dominate the observed bumps in this energy range.

\subsection{The contribution of an ensemble of sources to the GCR spectrum}

In order to estimate the total contribution of CSF systems to the
overall CR spectrum we have followed a model of CR diffusion in the
Galaxy \cite[see e.g.,][]{ginzburg64,Thoudam2014}. The analytical
solution of the transport equation was used to describe a steady CR
distribution in the Galactic plane, which  was assumed to be flat
and thin \cite[e.g.,][]{Thoudam2014}. Only primary accelerated
particles in the $10^{15}-10^{17}$ eV energy range were taken into
account.

A steady-state transport equation with diffusion and
interaction losses for CR nuclei in cylindrical
geometry can be written as
\begin{eqnarray}
\nabla (D \nabla N) + n v(p) \sigma(p) \delta (z) N(r, z,p) = - Q(p) \delta(z).
\label{diffusion}
\end{eqnarray}
Here $r$ and $z$ are the radial and vertical
coordinates, with z = 0 representing the galactic plane, $p$ is the momentum of the particle, $n$ is
the average number density, $v(p)$ is the velocity of a
particle, $D(p)$ is the diffusion coefficient, $\sigma(p)$ is the
inelastic collision cross-section,  and
$Q(p)$ is the source spectral energy
distribution shown in Fig.~\ref{particles}.

The solution of (\ref{diffusion}) can be written as
\begin{eqnarray}
N(z, p) =
\nu R
\int_0^{\infty} dk \frac{\sinh{[k(L-z)]}}{\sinh{(k L)}}\cdot
\frac{J_1(k R)}{B(p)}\cdot Q(p),
\label{solution}
\end{eqnarray}
%
%
where
\begin{eqnarray}
B(p)=2 D(p) \coth{(kL)} +n v(p) \sigma(p),
\end{eqnarray}
$J_1(x)$ is the Bessel function of index 1, and  $\nu$ = 2.5
sources/Myr/kpc$^2$ is the rate of SNe in clusters with CSF
systems in the Galaxy. The rate $\nu$ was assumed to be 10\% of the
total galactic SNe rate, which, in turn, is taken to be 3 SNe per
100 years. Thus we assume that there are $\sim 10$ CSF systems
accelerating CRs in the Milky Way at any given time. These data were
chosen to illustrate a possible contribution of the CSF systems to
the observed CR spectrum. More detailed observational information on
the galactic SNe in young massive star clusters is needed to justify
or reject the model assumptions.

\begin{figure}  
\resizebox{15cm}{11.5cm}{
\includegraphics[trim= 130 260 140 320,clip]{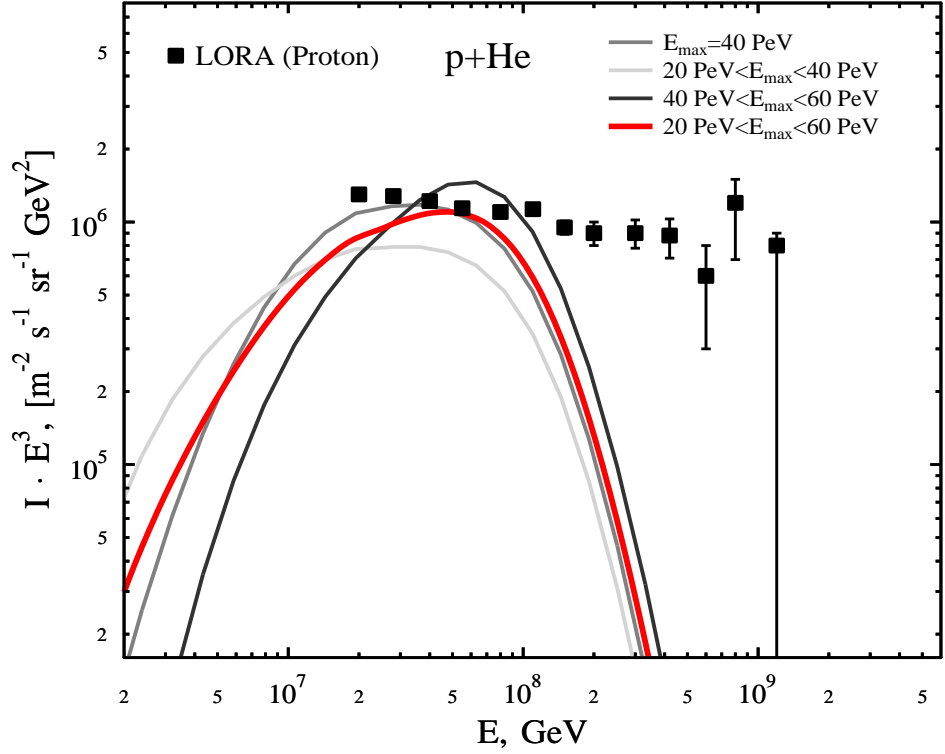}}
\caption{Energy spectra of the light-mass CR component measured by LOFAR
\citep[][]{Buitink2016} compared to non-linear model predictions for the CR flux from
galactic CSF systems.
The model CR spectra for a mixture of $50\%$ p and $50\%$ He, for
different values of $\Emax$  (from 20 to 60 PeV),
are shown as gray and red lines.
The model normalization is obtained by assuming that $10\%$ of all galactic SNRs are in CSF systems and that $30\%$ of the shock kinetic energy is converted into CRs.}
\label{LOFAR}
\end{figure}

In Fig.~\ref{LOFAR} we compare our galactic CSF model predictions to
LOFAR data as presented by \citet{Buitink2016}.
We assume that the maximum energy of CR protons produced by a single
source varies from $20$ to $60$\,PeV.
As discussed above, such energies can be reached within $300-400$ yr.

The best fit to the LOFAR data is achieved for a four-component
model of protons, helium, nitrogen, and iron nuclei
with a fraction of protons and helium of 0.8 
\citep{Buitink2016}. In order to illustrate the possible
contribution of the light component accelerated at CSF sites to the
overall CR spectrum in Fig.~\ref{LOFAR}, simulated spectra for a
mixture of $50\%$ proton and $50\%$ helium cumulatively produced by
galactic CSFs are shown for different distributions of $\Emax$.
The simulations assume the efficiency of energy conversion into CRs is $\eta = 0.3$ and with the above parameters, our results show that
a set of galactic CSF systems in massive stellar clusters can
dominate the observed CR flux in the energy range
$10^{16.5}-10^{17}$ eV. We propose this as the second galactic
component which is suggested by LOFAR measurements.

\section{Neutrino and $\gamma$-ray emission from supernovae in stellar clusters}
Previously  we have modeled CRs and neutrinos produced by colliding
shocks in the compact star cluster \WestI. Using parameters for the
presumed supernova progenitor of the magnetar \CXO, we concluded
that this galactic source could explain a fraction of the recently
detected {\sl IceCube} neutrinos \citep{Bykov2015}.

We now consider a supernova which produced the 650 year old
magnetar SGR 1806-20, which is a member of the massive stellar
cluster Cl*~1806-20 \citep{Fuchs1999, Corbel2004}.
The soft-gamma-repeater SGR 1806-20 is
known for its giant flare in 2004 and is one of the most interesting
and active soft gamma repeaters (SGRs). The age of the cluster
Cl*~1806-20 is estimated to be about 3-4~Myr and it contains many
massive stars including four Wolf-Rayet (WR) stars, five O-type
stars, and a luminous blue variable candidate LBV~1806-20
\citep{vanKerkwijk1995}.
The
supernova shock that must have been launched after the birth of
SGR 1806-20 has had enough time to cross the stellar cluster and may provide
the conditions required for efficient particle acceleration according to the CSF model
discussed above.

\begin{figure}  
\resizebox{12cm}{10.5cm}{
\includegraphics[trim= 0 0 0 0,clip]{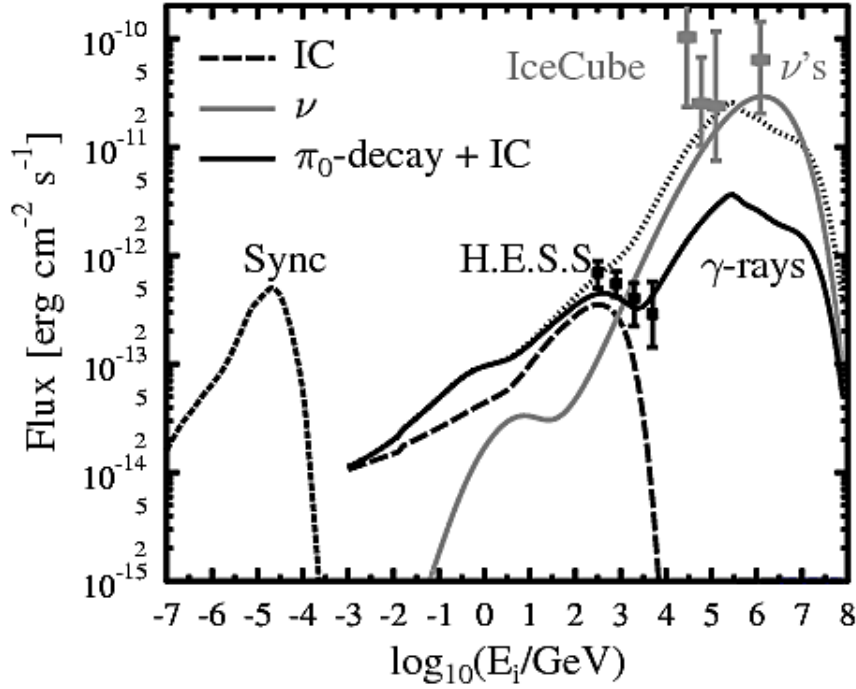}}
\caption{Model predictions of \gamray\ flux (black) and neutrino
flux (gray) from $pp$-interactions in a region around the CSF system
in the stellar cluster CL*~1806-20. Pair
production at the highest energy range  is included and we assume a distance of
$8.7$\,kpc \citep{Bibby2008}. A 15~pc radius of the $\gamma$-ray
emission source is consistent with the H.E.S.S. data. Dashed curves
show the inverse Compton (IC) and synchrotron emission from primary
and secondary (from $\pi^0$-decay) electrons accelerated directly in
this source. The ISRF near the Galactic Center used for the
calculation of the IC emission was adopted from GALPROP
\citep{Porter2008}. The extreme upward curvature in the neutrino
spectrum above $E=10$\,TeV reflects the transition from CR acceleration
at a single SNR shock for low-energy particles to the more efficient
acceleration of high-energy particles as they scatter back and forth
between the SNR shock and the cluster wind. The data points for the
H.E.S.S. source \citep{HESS1806}, and the energy flux corresponding
to four \Ice\ events in the vicinity of the source HESS~J1806-204
are shown. } \label{SpectrSGR1806}
\end{figure}

Given the similarity of the $\gamma$-ray source HESS J1808-204 with
previously detected H.E.S.S. sources in massive star associations
Cyg OB2, \WestI, and \WestII\  \citep{aharCygOB_02,hessWd1}, we have
applied the CSF particle acceleration model to the stellar cluster
Cl* 1806-20 taken to be at a distance of 8.7\,kpc.
In Fig.~\ref{SpectrSGR1806} we compare predicted $\gamma$-ray and
neutrino spectra with \Hess\ and \Ice\ observations. The predicted
spectra include emission  from CRs still trapped in the source as
well as those that have escaped from the accelerator and diffused
into the surrounding cloudy medium over the last 650\,yr. The
trapped and escaped CR protons and electrons used to generate
Fig.~\ref{SpectrSGR1806} are shown in Fig.~\ref{particles}.
The upward curvature in the neutrino spectrum above $\sim 10$\,TeV
reflects the transition from CR acceleration at the single SNR shock
to the more efficient acceleration of high-energy
particles as they scatter back and forth between the SNR shock and
the cluster wind.

The diffusion coefficient within $\sim$ 30 pc of the CSF accelerator
was taken to be $D=5\cdot10^{29}(E/\mathrm{PeV})^{0.33}$\,\csps\
which is broadly consistent with the ISM values as discussed by
\citet{Strong2007}. The number density in the cloudy medium in the
dense region around the accelerator was assumed to be $\sim
90$\,\pcc.  This value seems to be justified even if the cluster Cl*
1806-20 is located away from the Galactic center region because the
young cluster likely originated from a molecular cloud which should
have a much longer life time and may extend for $\sim$ 20-30 pc.

A possible association of Cl* 1806-20 and SGR 1806-20 with the
Galactic center region is still a matter of a dispute. To construct
the spectra in Figure \ref{SpectrSGR1806} we adopted a distance of
$8.7$\,kpc following the analysis by \citet{Bibby2008}. However,
\citet{Svirski2011} constrained the distance to SGR 1806-20 with a
lower limit of 9.4 kpc and an upper limit of 18.6 kpc (90 per cent
confidence). \citet{Tendulkar12}, who measured the proper motion of
SGR 1806-20, has adopted the distance to SGR 1806-20 to be 9 $\pm$ 2
kpc, which is consistent with all the measurements available.

\begin{figure}     
\resizebox{13cm}{10.5cm}{
\includegraphics{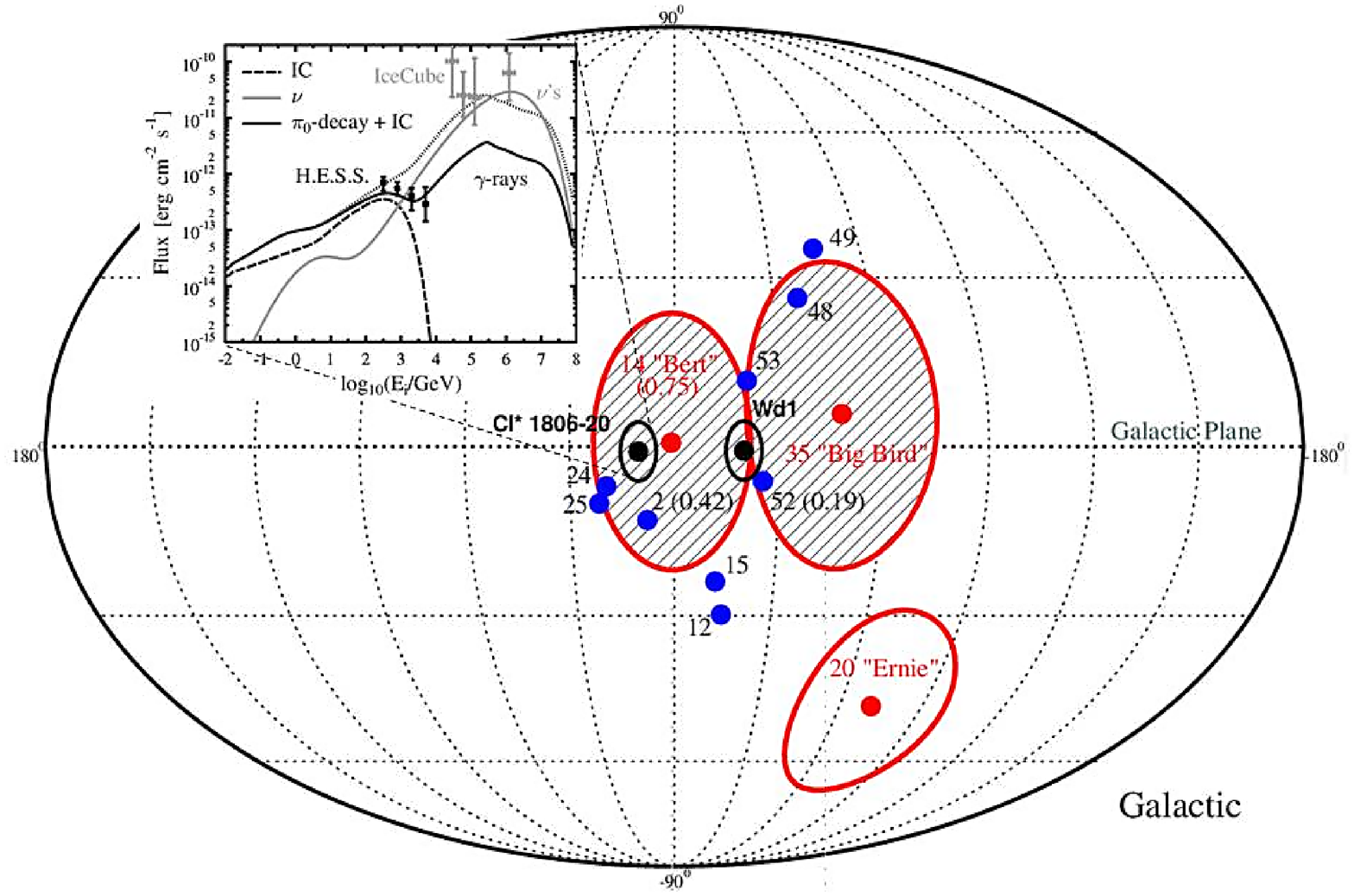}}
\caption{Map showing subset of PeV and subPeV \Ice\ events possibly
associated with the galactic stellar clusters Westerlund 1 (Wd1) and
Cl*1806-20.  Position uncertainty and median angular errors for the
PeV events (2-$\sigma$ red contours) was determined from 2D Gaussian
statistics \citep{Aartsen14}. The high-energy spectral insert for
the CSF system in Cl*1806-20 points to the position of the cluster
(see also fig.~\ref{SpectrSGR1806}).
Two PeV events 14 (``Bert")  and 35 (``Big Bird") are within
2$\sigma$ from both Wd1 and Cl* 1806-20. Positions of the nearest
Sub-Pev events, including one track event $�52$ are also shown. For
the events 2, 14, and 52 in brackets we indicate the probabilities
for these neutrinos to be  galactic \citep[according to the analyzes in][]{Denton2017}. It is interesting to note that these events
have the highest probabilities for the galactic origin among the
full set of \Ice\ neutrinos detected to date. } \label{map}
\end{figure}

 The \Ice\ data points in Fig.~\ref{SpectrSGR1806} show the neutrino energy flux
consistent with the position of Cl* 1806-20 at the angular
resolution of \Ice\ \citep{Aartsen14}. Thus, the CSF model applied
to HESS J1806-204 can explain a subset of the observed \Ice\
neutrinos in the inner Galaxy.
As we show in Fig.~\ref{map}, one of the PeV events (14, ``Bert")
from recent neutrino detections is within $1\sigma$ angular distance
from the position of Cl* 1806-20 as determined from $2D$ Gaussian
statistics and the median angular errors and positions given by
\citet{Aartsen14}.

Primary electrons accelerated in the CSF system and secondary
electrons from $\piZ$-decay are both contributing to the inverse
Compton (IC) spectrum in Fig.~\ref{SpectrSGR1806}. The ISRF field
near the Galactic Center needed for the simulation of the IC
spectrum was adopted from a GALPROP simulation \citep{Porter2008}.
Summing the lepton emission with the hadron emission leads to the
softening of the resulting spectrum in the  100~GeV -- 1~TeV energy
range, fitting the \Hess\ data. For completeness we also show in
Fig.~\ref{SpectrSGR1806} the \syn\ emission from primary and
secondary electrons accelerated in the CSF system.

For illustration in Fig.~\ref{map} we  show a galactic map for a
subset of neutrinos detected by \Ice\ in the vicinity of \WestI\ and
Cl*1806-20. Two-$\sigma$ contours (applying Gaussian statistics) are
shown for PeV events and one can see that two PeV events 14
(``Bert") and 35 (``Big Bird") are within 2$\sigma$ from both
\WestI\ and Cl* 1806-20.
We note that while no time  or space clumping has been confirmed  so
far, it has been reported  that not less than $25-30\%$ of the \Ice\
events should originate from the Galactic plane \citep{IceCube2017}.
In another analyzes, \citet{Denton2017}  pointed out that the
maximum likelihood function for the origin of neutrino events
corresponds to a $6.6\%$ portion of Galactic neutrinos. This
analyzes shows that among the full set of IceCube neutrinos,  events
2, 14 and 52 have the highest probabilities for being galactic with
0.42, 0,75 and 0.19 probabilities respectively.

\section{Conclusions}
The expanding young supernova shells in compact stellar clusters can
be very efficient CR accelerators to energies well
above the standard limits of diffusive shock acceleration in an
isolated SNR.
During a relatively short time of a few hundred years after the supernova explodes, the SNR shock can collide with the fast wind from a massive star, or with the cluster wind, and produce Fermi accelerated CRs to energies
above $10^{16}$\,eV.
The energy spectrum of protons in such a colliding shock flow accelerator is  hard,
approaching  the spectral shape $dN/dp \propto p^{-1}$ before a
break, providing a large energy flux in the high-energy end of the
spectrum.
While the strongest acceleration stage lasts for only a  few hundred
years, high-energy CRs escape the CSF system and diffuse through
the ambient matter producing $\gamma$-rays and neutrinos in
inelastic nuclear collisions over a much longer period. We have
modeled a galactic population of CSF sources and conclude CSF systems would
contribute a significant light-mass component to galactic CRs above
$\sim 10^{16}$\,eV. We compare our predictions to recent LOFAR
observations \citep{Buitink2016} which suggest a need for such a
component.
According to the analyses of \citet{Buitink2016},
the total fraction of light elements (p and He) in LOFAR
measurements lies between 0.38 and 0.98 for a significance level
$p>0.01$.
 The light component mix is likely to have both protons and He
with an excess of He nuclei. We have simulated a 50/50 mix of protons
and He, but a slight change in this ratio will not strongly affect our
result.

Considering neutrinos, the recent data release of the {\sl IceCube} Collaboration has
confirmed that no individual steady-state or transient source of
very high-energy neutrinos has been identified so far
\citep{IceCube2017}. An analysis of the data obtained during 6 years
suggests that AGN jets or GRBs are likely to be responsible
for only a small fraction of the observed neutrinos \citep{IceCube2017}.
Thus, there is a substantial need for alternative models of
efficient sources of very high-energy cosmic neutrinos.
We have proposed that CSF systems in galactic compact
clusters of massive stars may produce CRs and neutrinos efficiently
and that two such clusters, \WestI\ and Cl*1806-20, are spatially
consistent  with a fraction of the \Ice\ neutrinos. \textit{Events
2, 14, and 52 being within a 2-$\sigma$ distance from both \WestI\
and Cl*1806-20 have the highest probabilities for being galactic
with 0.42, 0.75, and 0.19 probabilities respectively.} Consistently
with the neutrino production, we have shown that the CRs in these
two clusters can account for the TeV \gamray\ emission detected by
H.E.S.S.

Recently, \citet{neronov16}  showed that the galactic
latitude distribution of the astrophysical neutrino events with
energies above 100 TeV detected with {\sl IceCube} is apparently
inconsistent with an isotropic distribution. This supports our conjecture that SNe in galactic compact
clusters, namely \WestI\ and Cl*1806-20, may explain $\gsim 10\%$ of the observed neutrino flux.
Moreover, compact clusters are numerous in starburst galaxies.
The starburst galaxies have been proposed  by
\citet[][]{loeb_waxman06} as important sources of extra-galactic
neutrinos and these are widely discussed in the context of the \Ice\
data \citep[see
also][]{starburst_neutrinos14,SFneutrinosJCAP15,b14,againstSF15}. We
believe it is possible that the colliding shock flow model we
propose is the main PeV CR generator in starburst galaxies and
therefore  responsible for a substantial fraction of extra-galactic
\Ice\ neutrinos as well.
%
%
%
The development
of the next generation instrument, {\it IceCube-Gen2}, due to its five
times better sensitivity for individual sources and not less than
ten times larger surface area, will allow  detailed
studies of high-energy cosmic neutrinos, identify
their powerful sources, and investigate the energy conversion
processes therein  \citep[][]{IceCubeGen2}.

\section{Acknowledgements}
We thank the anonymous referee for constructive comments. The
results of the work were obtained using computational resources of
Peter the Great Saint-Petersburg Polytechnic University
Supercomputing Center.


\begin{thebibliography}{52}
\ifx \bisbn   \undefined \def \bisbn  #1{ISBN #1}\fi \ifx \binits
\undefined \def \binits#1{#1} \fi \ifx \bauthor  \undefined \def
\bauthor#1{#1} \fi \ifx \bjtitle  \undefined \def
\bjtitle#1{\textrm{#1}}\fi \ifx \batitle  \undefined \def
\batitle#1{#1} \fi \ifx \bctitle  \undefined \def \bctitle#1{#1} \fi
\ifx \bvolume  \undefined \def \bvolume#1{\textbf{#1}}\fi \ifx
\byear  \undefined \def \byear#1{#1} \fi \ifx \bissue  \undefined
\def \bissue#1{#1} \fi \ifx \bfpage  \undefined \def \bfpage#1{#1}
\fi \ifx \blpage  \undefined \def \blpage #1{#1} \fi \ifx \burl
\undefined \def \burl#1{#1} \fi \ifx \doiurl  \undefined \def
\doiurl#1{#1} \fi \ifx \betal  \undefined \def \betal{et al.} \fi
\ifx \binstitute  \undefined \def \binstitute#1{#1} \fi \ifx
\beditor  \undefined \def \beditor#1{#1} \fi \ifx \bpublisher
\undefined \def \bpublisher#1{#1} \fi \ifx \bbtitle  \undefined \def
\bbtitle#1{\textit{#1}} \fi \ifx \bedition  \undefined \def
\bedition#1{#1} \fi \ifx \bseriesno  \undefined \def
\bseriesno#1{#1} \fi \ifx \blocation  \undefined \def
\blocation#1{#1} \fi \ifx \bsertitle  \undefined \def
\bsertitle#1{#1} \fi \ifx \bsnm \undefined \def \bsnm#1{#1} \fi \ifx
\bsuffix \undefined \def \bsuffix#1{#1} \fi \ifx \bparticle
\undefined \def \bparticle#1{#1} \fi \ifx \barticle \undefined \def
\barticle#1{#1} \fi \ifx \botherref \undefined \def \botherref
#1{#1} \fi \ifx \url \undefined \def \url#1{#1} \fi \ifx \bchapter
\undefined \def \bchapter#1{#1} \fi \ifx \bbook \undefined \def
\bbook#1{#1} \fi \ifx \bcomment \undefined \def \bcomment#1{#1} \fi
\ifx \oauthor \undefined \def \oauthor#1{#1} \fi \ifx
\citeauthoryear \undefined \def \citeauthoryear#1{#1} \fi \ifx
\texttildelow  \undefined \def \texttildelow{\symbol{126}} \fi
\def \endbibitem {}
\ifx \bconflocation  \undefined \def \bconflocation#1{#1} \fi

\bibitem[\protect\citeauthoryear{{Aartsen} et~al.}{2013}]{aartsen2013}
\begin{barticle}
\bauthor{\binits{M.G.} \bsnm{{Aartsen}}}, \bauthor{\binits{R.}
\bsnm{{Abbasi}}}, \bauthor{\binits{Y.} \bsnm{{Abdou}}}, \betal,
\batitle{{Measurement of the cosmic ray energy spectrum with
IceTop-73}}. \bjtitle{\prd}, \bvolume{88}(\bissue{4}),
\bfpage{042004} (\byear{2013}).
\end{barticle}
\endbibitem

\bibitem[\protect\citeauthoryear{{Aartsen} et~al.}{2014}]{Aartsen14}
\begin{barticle}
\bauthor{\binits{M.G.} \bsnm{{Aartsen}}}, \bauthor{\binits{M.}
\bsnm{{Ackermann}}}, \bauthor{\binits{J.} \bsnm{{Adams}}}, \betal,
\batitle{{Observation of High-Energy Astrophysical Neutrinos in
Three Years of
  IceCube Data}}.
\bjtitle{\PRL}, \bvolume{113}(\bissue{10}),
\bfpage{101101} (\byear{2014}).
\end{barticle}
\endbibitem

\bibitem[\protect\citeauthoryear{{Abdalla} et~al.}{2016}]{HESS1806}
\begin{botherref}
\oauthor{\binits{H.} \bsnm{{Abdalla}}}, \oauthor{\binits{A.}
\bsnm{{Abramowski}}}, \oauthor{\binits{F.} \bsnm{{Aharonian}}}, et
al., {Extended VHE gamma-ray emission towards SGR1806-20,
LBV1806-20, and stellar
  cluster Cl*1806-20}.
ArXiv e-prints (2016)
\end{botherref}
\endbibitem

\bibitem[\protect\citeauthoryear{{Abramowski} et~al.}{2012}]{hessWd1}
\begin{barticle}
\bauthor{\binits{A.} \bsnm{{Abramowski}}}, \bauthor{\binits{F.}
\bsnm{{Acero}}}, \bauthor{\binits{F.} \bsnm{{Aharonian}}}, \betal,
\batitle{{Discovery of extended VHE {$\gamma$}-ray emission from the
vicinity
  of the young massive stellar cluster Westerlund 1}}.
\bjtitle{\aap}, \bvolume{537}, \bfpage{114} (\byear{2012}).
\end{barticle}
\endbibitem

\bibitem[\protect\citeauthoryear{{Ackermann} et~al.}{2011}]{fermiSB11}
\begin{barticle}
\bauthor{\binits{M.} \bsnm{{Ackermann}}}, \bauthor{\binits{M.}
\bsnm{{Ajello}}}, \bauthor{\binits{A.e.} \bsnm{{Allafort}}},
\batitle{{A Cocoon of Freshly Accelerated Cosmic Rays Detected by
Fermi in the
  Cygnus Superbubble}}.
\bjtitle{Science}, \bvolume{334}, \bfpage{1103} (\byear{2011}).
\end{barticle}
\endbibitem

\bibitem[\protect\citeauthoryear{{Aharonian} et~al.}{2002}]{aharCygOB_02}
\begin{barticle}
\bauthor{\binits{F.} \bsnm{{Aharonian}}}, \bauthor{\binits{A.}
\bsnm{{Akhperjanian}}}, \bauthor{\binits{M.} \bsnm{{Beilicke}}},
\betal, \batitle{{An unidentified TeV source in the vicinity of
Cygnus OB2}}. \bjtitle{\aap}, \bvolume{393}, \bfpage{37}--\blpage{40}
(\byear{2002}). 
\end{barticle}
\endbibitem

\bibitem[\protect\citeauthoryear{{Amato}}{2014}]{amato14}
\begin{barticle}
\bauthor{\binits{E.} \bsnm{{Amato}}}, \batitle{{The origin of
galactic cosmic rays}}. \bjtitle{International Journal of Modern
Physics D} \bvolume{23}, \bfpage{30013} (\byear{2014}).
\end{barticle}
\endbibitem

\bibitem[\protect\citeauthoryear{{Amenomori} et~al.}{2008}]{amenomori2008}
\begin{barticle}
\bauthor{\binits{M.} \bsnm{{Amenomori}}}, \bauthor{\binits{X.J.}
\bsnm{{Bi}}}, \bauthor{\binits{D.} \bsnm{{Chen}}}, \betal,
\batitle{{The All-Particle Spectrum of Primary Cosmic Rays in the
Wide Energy
  Range from 10$^{14}$ to 10$^{17}$ eV Observed with the Tibet-III Air-Shower
  Array}}.
\bjtitle{\apj}, \bvolume{678}, \bfpage{1165}--\blpage{1179}
(\byear{2008}). 
\end{barticle}
\endbibitem

\bibitem[\protect\citeauthoryear{{Apel} et~al.}{2009}]{apel2009}
\begin{barticle}
\bauthor{\binits{W.D.} \bsnm{{Apel}}}, \bauthor{\binits{J.C.}
\bsnm{{Arteaga}}}, \bauthor{\binits{A.F.} \bsnm{{Badea}}}, \betal,
\batitle{{Energy spectra of elemental groups of cosmic rays: Update
on the
  KASCADE unfolding analysis}}.
\bjtitle{\APP}, \bvolume{31},
\bfpage{86}--\blpage{91} (\byear{2009}).
\end{barticle}
\endbibitem

\bibitem[\protect\citeauthoryear{{Apel} et~al.}{2013}]{KASCADE13}
\begin{bchapter}
\bauthor{\binits{W.D.} \bsnm{{Apel}}}, \bauthor{\binits{J.C.}
\bsnm{{Arteaga-Velazquez}}}, \bauthor{\binits{K.} \bsnm{{Bekk}}},
\betal, \bctitle{{Separation of the light and heavy mass groups of
10$^{16}$ -
  10$^{18}$ eV cosmic rays by studying the ratio muon size to shower size of
  KASCADE-Grande data}},
in \bbtitle{Journal of Physics Conference Series}.
\bsertitle{Journal of Physics Conference Series}, vol.
\bseriesno{409}, \byear{2013}, p. \bfpage{012095}.
\end{bchapter}
\endbibitem

\bibitem[\protect\citeauthoryear{{Berezhnev} et~al.}{2012a}]{berezhnevea13}
\begin{barticle}
\bauthor{\binits{S.F.} \bsnm{{Berezhnev}}}, \bauthor{\binits{D.}
\bsnm{{Besson}}}, \bauthor{\binits{N.M.} \bsnm{{Budnev}}}, \betal,
\batitle{{The Tunka-133 EAS Cherenkov light array: Status of 2011}}.
\bjtitle{Nuclear Instruments and Methods in Physics Research A}
\bvolume{692}, \bfpage{98}--\blpage{105} (\byear{2012}a).
\end{barticle}
\endbibitem

\bibitem[\protect\citeauthoryear{{Bechtol} et~al.}{2017}]{againstSF15}
\begin{barticle}
\bauthor{\binits{K.} \bsnm{{Bechtol}}}, \bauthor{\binits{M.}
\bsnm{{Ahlers}}}, \bauthor{\binits{M.} \bsnm{{Di Mauro}}}, \betal,
\batitle{{Evidence against Star-forming Galaxies as the Dominant
Source of
  Icecube Neutrinos}}.
\bjtitle{\apj}, \bvolume{836}, \bfpage{47} (\byear{2017}).
\end{barticle}
\endbibitem



\bibitem[\protect\citeauthoryear{{Berezhnev} et~al.}{2012b}]{Prosin2014}
\begin{barticle}
\bauthor{\binits{S.F.} \bsnm{{Berezhnev}}}, \bauthor{\binits{D.}
\bsnm{{Besson}}}, \bauthor{\binits{N.M.} \bsnm{{Budnev}}}, \betal,
\batitle{{The Tunka-133 EAS Cherenkov light array: Status of 2011}}.
\bjtitle{Nuclear Instruments and Methods in Physics Research A}
\bvolume{692}, \bfpage{98}--\blpage{105} (\byear{2012}b).
\end{barticle}
\endbibitem

\bibitem[\protect\citeauthoryear{{Bibby} et~al.}{2008}]{Bibby2008}
\begin{barticle}
\bauthor{\binits{J.L.} \bsnm{{Bibby}}}, \bauthor{\binits{P.A.}
\bsnm{{Crowther}}}, \bauthor{\binits{J.P.} \bsnm{{Furness}}},
\betal, \batitle{{A downward revision to the distance of the 1806-20
cluster and
  associated magnetar from Gemini Near-Infrared Spectroscopy}}.
\bjtitle{MNRAS},
\bvolume{386}, \bfpage{23}--\blpage{27} (\byear{2008}).
\end{barticle}
\endbibitem

\bibitem[\protect\citeauthoryear{{Blandford} et~al.}{2014}]{blandford14}
\begin{barticle}
\bauthor{\binits{R.} \bsnm{{Blandford}}}, \bauthor{\binits{P.}
\bsnm{{Simeon}}}, \bauthor{\binits{Y.} \bsnm{{Yuan}}},
\batitle{{Cosmic Ray Origins: An Introduction}}. \bjtitle{Nuclear
Physics B Proceedings Supplements} \bvolume{256},
\bfpage{9}--\blpage{22} (\byear{2014}).
\end{barticle}
\endbibitem

\bibitem[\protect\citeauthoryear{{Blasi}}{2013}]{blasiAARv13}
\begin{barticle}
\bauthor{\binits{P.} \bsnm{{Blasi}}}, \batitle{{The origin of
galactic cosmic rays}}. \bjtitle{\aapr}, \bvolume{21}, \bfpage{70}
(\byear{2013}). 
\end{barticle}
\endbibitem

\bibitem[\protect\citeauthoryear{{Budnev} et~al.}{2013}]{Budnev2013}
\begin{barticle}
\bauthor{\binits{N.} \bsnm{{Budnev}}}, \bauthor{\binits{D.}
\bsnm{{Chernov}}}, \bauthor{\binits{O.} \bsnm{{Gress}}}, \betal,
\batitle{{Tunka-25 Air Shower Cherenkov array: The main results}}.
\bjtitle{\APP}, \bvolume{50},
\bfpage{18}--\blpage{25} (\byear{2013}).
\end{barticle}
\endbibitem

\bibitem[\protect\citeauthoryear{{Buitink} et~al.}{2016}]{Buitink2016}
\begin{barticle}
\bauthor{\binits{S.} \bsnm{{Buitink}}}, \bauthor{\binits{A.}
\bsnm{{Corstanje}}}, \bauthor{\binits{H.} \bsnm{{Falcke}}}, \betal,
\batitle{{A large light-mass component of cosmic rays at
10$^{17}$-10$^{17.5}$
  electronvolts from radio observations}},
\bjtitle{\nat}, \bvolume{531}, \bfpage{70}--\blpage{73}
(\byear{2016}). 
\end{barticle}
\endbibitem

\bibitem[\protect\citeauthoryear{{Bykov}}{2001}]{b01}
\begin{barticle}
\bauthor{\binits{A.M.} \bsnm{{Bykov}}}, \batitle{{Particle
Acceleration and Nonthermal Phenomena in Superbubbles}}.
\bjtitle{\ssr}, \bvolume{99},
\bfpage{317}--\blpage{326} (\byear{2001})
\end{barticle}
\endbibitem

\bibitem[\protect\citeauthoryear{{Bykov}}{2014}]{b14}
\begin{barticle}
\bauthor{\binits{A.M.} \bsnm{{Bykov}}}, \batitle{{Nonthermal
particles and photons in starburst regions and
  superbubbles}}.
\bjtitle{\aapr}, \bvolume{22}, \bfpage{77} (\byear{2014}).
\end{barticle}
\endbibitem

\bibitem[\protect\citeauthoryear{{Bykov} and {Toptygin}}{2001}]{bt01}
\begin{barticle}
\bauthor{\binits{A.M.} \bsnm{{Bykov}}}, \bauthor{\binits{I.N.}
\bsnm{{Toptygin}}}, \batitle{{A Model of Particle Acceleration to
High Energies by Multiple
  Supernova Explosions in OB Associations}}.
\bjtitle{Astronomy Letters}, \bvolume{27}, \bfpage{625}--\blpage{633}
(\byear{2001}). 
\end{barticle}
\endbibitem

\bibitem[\protect\citeauthoryear{{Bykov} et~al.}{2013}]{MNRAS_BGO13}
\begin{barticle}
\bauthor{\binits{A.M.} \bsnm{{Bykov}}}, \bauthor{\binits{P.E.}
\bsnm{{Gladilin}}}, \bauthor{\binits{S.M.} \bsnm{{Osipov}}},
\batitle{{Non-linear model of particle acceleration at colliding
shock flows}}. \bjtitle{MNRAS}, \bvolume{429},
\bfpage{2755}--\blpage{2762} (\byear{2013}).
\end{barticle}
\endbibitem

\bibitem[\protect\citeauthoryear{{Bykov} et~al.}{2015}]{Bykov2015}
\begin{barticle}
\bauthor{\binits{A.M.} \bsnm{{Bykov}}}, \bauthor{\binits{D.C.}
\bsnm{{Ellison}}}, \bauthor{\binits{P.E.} \bsnm{{Gladilin}}},
\betal, \batitle{{Ultrahard spectra of PeV neutrinos from supernovae
in compact star
  clusters}}.
\bjtitle{\mnras}, \bvolume{453}, \bfpage{113}--\blpage{121}
(\byear{2015}). 
\end{barticle}
\endbibitem

\bibitem[\protect\citeauthoryear{{Clark} et~al.}{2008}]{Clark2008}
\begin{barticle}
\bauthor{\binits{J.S.} \bsnm{{Clark}}}, \bauthor{\binits{M.P.}
\bsnm{{Muno}}}, \bauthor{\binits{I.} \bsnm{{Negueruela}}}, \betal,
\batitle{{Unveiling the X-ray point source population of the Young
Massive
  Cluster Westerlund 1}}.
\bjtitle{\aap}, \bvolume{477}, \bfpage{147}--\blpage{163}
(\byear{2008}). 
\end{barticle}
\endbibitem

\bibitem[\protect\citeauthoryear{{Corbel} and {Eikenberry}}{2004}]{Corbel2004}
\begin{barticle}
\bauthor{\binits{S.} \bsnm{{Corbel}}}, \bauthor{\binits{S.S.}
\bsnm{{Eikenberry}}}, \batitle{{The connection between W31, SGR
1806-20, and LBV 1806-20: Distance,
  extinction, and structure}}.
\bjtitle{\aap}, \bvolume{419}, \bfpage{191}--\blpage{201}
(\byear{2004}). 
\end{barticle}
\endbibitem

\bibitem[\protect\citeauthoryear{{Denton} et~al.}{2017}]{Denton2017}
\begin{botherref}
\oauthor{\binits{P.B.} \bsnm{{Denton}}}, \oauthor{\binits{D.}
\bsnm{{Marfatia}}}, \oauthor{\binits{T.J.} \bsnm{{Weiler}}}, {The
Galactic Contribution to IceCube's Astrophysical Neutrino Flux}.
ArXiv e-prints (2017)
\end{botherref}
\endbibitem


\bibitem[\protect\citeauthoryear{{Ellison} et~al.}{2013b}]{MCtransrel13}
\begin{barticle}
\bauthor{\binits{D.C.} \bsnm{{Ellison}}}, \bauthor{\binits{D.C.}
\bsnm{{Warren}}}, \bauthor{\binits{A.M.} \bsnm{{Bykov}}},
\batitle{{Monte Carlo Simulations of Nonlinear Particle Acceleration
in
  Parallel Trans-relativistic Shocks}}.
\bjtitle{\apj}, \bvolume{776}, \bfpage{46} (\byear{2013}).
\end{barticle}
\endbibitem

\bibitem[\protect\citeauthoryear{{Emig} et~al.}{2015}]{SFneutrinosJCAP15}
\begin{barticle}
\bauthor{\binits{K.} \bsnm{{Emig}}}, \bauthor{\binits{C.}
\bsnm{{Lunardini}}}, \bauthor{\binits{R.} \bsnm{{Windhorst}}},
\batitle{{Do high energy astrophysical neutrinos trace star
formation?}} \bjtitle{\jcap}, \bvolume{12}, \bfpage{029}
(\byear{2015}). 
\end{barticle}
\endbibitem



\bibitem[\protect\citeauthoryear{{Ferrand} and {Marcowith}}{2010}]{fermarcow10}
\begin{barticle}
\bauthor{\binits{G.} \bsnm{{Ferrand}}}, \bauthor{\binits{A.}
\bsnm{{Marcowith}}}, \batitle{{On the shape of the spectrum of
cosmic rays accelerated inside
  superbubbles}}.
\bjtitle{\aap}, \bvolume{510}, \bfpage{101} (\byear{2010}).
\end{barticle}
\endbibitem

\bibitem[\protect\citeauthoryear{{Fuchs} et~al.}{2000}]{Fuchs1999}
\begin{barticle}
\bauthor{\binits{Y.} \bsnm{{Fuchs}}}, \bauthor{\binits{F.}
\bsnm{{Mirabel}}}, \bauthor{\binits{S.} \bsnm{{Chaty}}}, \betal,
\batitle{{ISO observations of the environment of the soft gamma-ray
repeater
  SGR 1806-20}}.
\bjtitle{Nuclear Physics B Proc. Suppl.}, \bvolume{80},
\bfpage{11} (\byear{2000})
\end{barticle}
\endbibitem

\bibitem[\protect\citeauthoryear{{Garyaka} et~al.}{2013}]{Garyaka13}
\begin{bchapter}
\bauthor{\binits{A.} \bsnm{{Garyaka}}}, \bauthor{\binits{R.}
\bsnm{{Martirosov}}}, \bauthor{\binits{S.} \bsnm{{Ter-Antonyan}}},
\betal, \bctitle{{Fine structure of all-particle energy spectrum in
the knee region}}, in \bbtitle{Journal of Physics Conference
Series}. \bsertitle{Journal of Physics Conference Series}, vol.
\bseriesno{409}, \byear{2013}, p. \bfpage{012081}.
\end{bchapter}
\endbibitem

\bibitem[\protect\citeauthoryear{{Ginzburg} and {Syrovatskii}}{1964}]{ginzburg64}
\begin{botherref}
\oauthor{\binits{V.L.} \bsnm{{Ginzburg}}}, \oauthor{\binits{S.I.}
\bsnm{{Syrovatskii}}}, \textit{{The Origin of Cosmic Rays}} (New
York: Macmillan) (1964). 
\end{botherref}
\endbibitem

\bibitem[\protect\citeauthoryear{{Gladilin} et~al.}{2014}]{gladilin14}
\begin{barticle}
\bauthor{\binits{P.E.} \bsnm{{Gladilin}}}, \bauthor{\binits{A.M.}
\bsnm{{Bykov}}}, \bauthor{\binits{S.M.} \bsnm{{Osipov}}},
\batitle{{Maximal energies of the particles accelerated by the
system of
  converging magnetohydrodynamic flows}}.
\bjtitle{Journal of Physics Conference Series},
\bvolume{572}(\bissue{1}), \bfpage{012003} (\byear{2014}).
\end{barticle}
\endbibitem

\bibitem[\protect\citeauthoryear{{Hillas}}{2005}]{hillas05}
\begin{barticle}
\bauthor{\binits{A.M.} \bsnm{{Hillas}}}, \batitle{{TOPICAL REVIEW:
Can diffusive shock acceleration in supernova
  remnants account for high-energy galactic cosmic rays?}}
\bjtitle{Journal of Physics G Nuclear Physics}, \bvolume{31},
\bfpage{95} (\byear{2005}). 
\end{barticle}
\endbibitem

\bibitem[\protect\citeauthoryear{{IceCube Collaboration}
  et~al.}{2017}]{IceCube2017}
\begin{botherref}
\oauthor{\bsnm{{IceCube Collaboration}}}, \oauthor{\binits{M.G.}
\bsnm{{Aartsen}}}, \oauthor{\binits{M.} \bsnm{{Ackermann}}}, et al.,
{Neutrinos and Cosmic Rays Observed by IceCube}. ArXiv e-prints
(2017)
\end{botherref}
\endbibitem

\bibitem[\protect\citeauthoryear{{IceCube-Gen2 Collaboration}
  et~al.}{2014}]{IceCubeGen2}
\begin{botherref}
\oauthor{\bsnm{{IceCube-Gen2 Collaboration}}}, \oauthor{\bsnm{{:}}},
\oauthor{\binits{M.G.} \bsnm{{Aartsen}}}, et al., {IceCube-Gen2: A
Vision for the Future of Neutrino Astronomy in Antarctica}. ArXiv
e-prints (2014)
\end{botherref}
\endbibitem

\bibitem[\protect\citeauthoryear{{Jokipii} and
  {Morfill}}{1987}]{1987ApJ...312..170J}
\begin{barticle}
\bauthor{\binits{J.R.} \bsnm{{Jokipii}}}, \bauthor{\binits{G.}
\bsnm{{Morfill}}}, \batitle{{Ultra-high-energy cosmic rays in a
galactic wind and its termination
  shock}}.
\bjtitle{\apj}, \bvolume{312}, \bfpage{170}--\blpage{177}
(\byear{1987}). 
\end{barticle}
\endbibitem

\bibitem[\protect\citeauthoryear{{Katsuta} et~al.}{2017}]{G25}
\begin{barticle}
\bauthor{\binits{J.} \bsnm{{Katsuta}}}, \bauthor{\binits{Y.}
\bsnm{{Uchiyama}}}, \bauthor{\binits{S.} \bsnm{{Funk}}},
\batitle{{Extended Gamma-ray Emission from the G25.0+0.0 Region: A
Star Forming
  Region Powered by the Newly Found OB Association?}}
\bjtitle{\apj}, \bvolume{839}, \bfpage{129} (\byear{2017}).
\end{barticle}
\endbibitem

\bibitem[\protect\citeauthoryear{{Loeb} and {Waxman}}{2006}]{loeb_waxman06}
\begin{barticle}
\bauthor{\binits{A.} \bsnm{{Loeb}}}, \bauthor{\binits{E.}
\bsnm{{Waxman}}}, \batitle{{The cumulative background of high energy
neutrinos from starburst
  galaxies}}.
\bjtitle{\jcap}, \bvolume{5}, \bfpage{003} (\byear{2006}).
\end{barticle}
\endbibitem


\bibitem[\protect\citeauthoryear{{Neronov} and {Semikoz}}{2016}]{neronov16}
\begin{barticle}
\bauthor{\binits{A.} \bsnm{{Neronov}}}, \bauthor{\binits{D.}
\bsnm{{Semikoz}}}, \batitle{{Evidence the Galactic contribution to
the IceCube astrophysical
  neutrino flux}}.
\bjtitle{\APP}, \bvolume{75},
\bfpage{60}--\blpage{63} (\byear{2016}).
\end{barticle}
\endbibitem

\bibitem[\protect\citeauthoryear{{Ohm}}{2016}]{2016CRPhy..17..585O}
\begin{barticle}
\bauthor{\binits{S.} \bsnm{{Ohm}}}, \batitle{{Starburst galaxies as
seen by gamma-ray telescopes}}. \bjtitle{Comptes Rendus Physique},
\bvolume{17}, \bfpage{585}--\blpage{593} (\byear{2016}).
\end{barticle}
\endbibitem

\bibitem[\protect\citeauthoryear{{Parizot} et~al.}{2004}]{parizot04}
\begin{barticle}
\bauthor{\binits{E.} \bsnm{{Parizot}}}, \bauthor{\binits{A.}
\bsnm{{Marcowith}}}, \bauthor{\binits{E.} \bsnm{{van der Swaluw}}},
\betal, \batitle{{Superbubbles and energetic particles in the
Galaxy. I. Collective
  effects of particle acceleration}}.
\bjtitle{\aap}, \bvolume{424}, \bfpage{747}--\blpage{760}
(\byear{2004}). 
\end{barticle}
\endbibitem

\bibitem[\protect\citeauthoryear{{Park} et~al.}{2015}]{Park2015}
\begin{barticle}
\bauthor{\binits{J.} \bsnm{{Park}}}, \bauthor{\binits{D.}
\bsnm{{Caprioli}}}, \bauthor{\binits{A.} \bsnm{{Spitkovsky}}},
\batitle{{Simultaneous Acceleration of Protons and Electrons at
Nonrelativistic
  Quasiparallel Collisionless Shocks}}.
\bjtitle{\prl}, \bvolume{114}(\bissue{8}),
\bfpage{085003} (\byear{2015}).
\end{barticle}
\endbibitem

\bibitem[\protect\citeauthoryear{{Portegies Zwart}
  et~al.}{2010}]{massive_star_cluster_2010ARA&A}
\begin{barticle}
\bauthor{\binits{S.F.} \bsnm{{Portegies Zwart}}},
\bauthor{\binits{S.L.W.} \bsnm{{McMillan}}}, \bauthor{\binits{M.}
\bsnm{{Gieles}}}, \batitle{{Young Massive Star Clusters}}.
\bjtitle{\araa}, \bvolume{48}, \bfpage{431}--\blpage{493}
(\byear{2010}). 
\end{barticle}
\endbibitem

\bibitem[\protect\citeauthoryear{{Porter} et~al.}{2008}]{Porter2008}
\begin{barticle}
\bauthor{\binits{T.A.} \bsnm{{Porter}}}, \bauthor{\binits{I.V.}
\bsnm{{Moskalenko}}}, \bauthor{\binits{A.W.} \bsnm{{Strong}}},
\betal, \batitle{{Inverse Compton Origin of the Hard X-Ray and Soft
Gamma-Ray Emission
  from the Galactic Ridge}}.
\bjtitle{\apj}, \bvolume{682}, \bfpage{400}--\blpage{407}
(\byear{2008}). 
\end{barticle}
\endbibitem

\bibitem[\protect\citeauthoryear{{Ptuskin} et~al.}{2010}]{Ptu10}
\begin{barticle}
\bauthor{\binits{V.} \bsnm{{Ptuskin}}}, \bauthor{\binits{V.}
\bsnm{{Zirakashvili}}}, \bauthor{\binits{E.-S.} \bsnm{{Seo}}},
\batitle{{Spectrum of Galactic Cosmic Rays Accelerated in Supernova
Remnants}}. \bjtitle{\apj}, \bvolume{718}, \bfpage{31}--\blpage{36}
(\byear{2010}). 
\end{barticle}
\endbibitem

\bibitem[\protect\citeauthoryear{{Strong} et~al.}{2007}]{Strong2007}
\begin{barticle}
\bauthor{\binits{A.W.} \bsnm{{Strong}}}, \bauthor{\binits{I.V.}
\bsnm{{Moskalenko}}}, \bauthor{\binits{V.S.} \bsnm{{Ptuskin}}},
\batitle{{Cosmic-Ray Propagation and Interactions in the Galaxy}}.
\bjtitle{Annual Review of Nuclear and Particle Science},
\bvolume{57}, \bfpage{285}--\blpage{327} (\byear{2007}).
\end{barticle}
\endbibitem

\bibitem[\protect\citeauthoryear{{Sveshnikova}}{2003}]{2003A&A...409..799S}
\begin{barticle}
\bauthor{\binits{L.G.} \bsnm{{Sveshnikova}}}, \batitle{{The knee in
the Galactic cosmic ray spectrum and variety in
  Supernovae}}.
\bjtitle{\aap}, \bvolume{409}, \bfpage{799}--\blpage{807}
(\byear{2003}). 
\end{barticle}
\endbibitem

\bibitem[\protect\citeauthoryear{{Sveshnikova} et~al.}{2013}]{Sveshnikova2013}
\begin{barticle}
\bauthor{\binits{L.G.} \bsnm{{Sveshnikova}}}, \bauthor{\binits{O.N.}
\bsnm{{Strelnikova}}}, \bauthor{\binits{V.S.} \bsnm{{Ptuskin}}},
\batitle{{Spectrum and anisotropy of cosmic rays at TeV-PeV-energies
and
  contribution of nearby sources}}.
\bjtitle{\APP}, \bvolume{50},
\bfpage{33}--\blpage{46} (\byear{2013}).
\end{barticle}
\endbibitem

\bibitem[\protect\citeauthoryear{{Svirski} et~al.}{2011}]{Svirski2011}
\begin{barticle}
\bauthor{\binits{G.} \bsnm{{Svirski}}}, \bauthor{\binits{E.}
\bsnm{{Nakar}}}, \bauthor{\binits{E.O.} \bsnm{{Ofek}}},
\batitle{{SGR 1806-20 distance and dust properties in molecular
clouds by
  analysis of flare X-ray echoes}}.
\bjtitle{\mnras}, \bvolume{415}, \bfpage{2485}--\blpage{2494}
(\byear{2011}). 
\end{barticle}
\endbibitem

\bibitem[\protect\citeauthoryear{{Tamborra}
  et~al.}{2014}]{starburst_neutrinos14}
\begin{barticle}
\bauthor{\binits{I.} \bsnm{{Tamborra}}}, \bauthor{\binits{S.}
\bsnm{{Ando}}}, \bauthor{\binits{K.} \bsnm{{Murase}}},
\batitle{{Star-forming galaxies as the origin of diffuse high-energy
  backgrounds: Gamma-ray and neutrino connections, and implications for
  starburst history}}.
\bjtitle{Journal of Cosmology and Astroparticle Physics},
\bvolume{2014}(\bissue{09}), \bfpage{043} (\byear{2014})
\end{barticle}
\endbibitem


\bibitem[\protect\citeauthoryear{{Tendulkar} et~al.}{2012}]{Tendulkar12}
\begin{barticle}
\bauthor{\binits{S.P.} \bsnm{{Tendulkar}}}, \bauthor{\binits{P.B.}
\bsnm{{Cameron}}}, \bauthor{\binits{S.R.} \bsnm{{Kulkarni}}},
\batitle{{Proper Motions and Origins of SGR 1806-20 and SGR
1900+14}}. \bjtitle{\apj}, \bvolume{761}, \bfpage{76} (\byear{2012}).
\end{barticle}
\endbibitem

\bibitem[\protect\citeauthoryear{{Thoudam} and
  {H{\"o}randel}}{2014}]{Thoudam2014}
\begin{barticle}
\bauthor{\binits{S.} \bsnm{{Thoudam}}}, \bauthor{\binits{J.R.}
\bsnm{{H{\"o}randel}}}, \batitle{{GeV-TeV cosmic-ray spectral
anomaly as due to reacceleration by weak
  shocks in the Galaxy}}.
\bjtitle{\aap}, \bvolume{567}, \bfpage{33} (\byear{2014}).
\end{barticle}
\endbibitem

\bibitem[\protect\citeauthoryear{{van Kerkwijk} and
  {Kulkarni}}{1995}]{vanKerkwijk1995}
\begin{barticle}
\bauthor{\binits{M.H.} \bsnm{{van Kerkwijk}}},
\bauthor{\binits{S.R.} \bsnm{{Kulkarni}}}, \batitle{{Spectroscopy of
the White Dwarf Companions of PSR 0655+64 and PSR
  0820+02}}.
\bjtitle{\apjl}, \bvolume{454}, \bfpage{L141} (\byear{1995}).
\end{barticle}
\endbibitem

\bibitem[\protect\citeauthoryear{{Voelk} and
  {Biermann}}{1988}]{1988ApJ...333L..65V}
\begin{barticle}
\bauthor{\binits{H.J.} \bsnm{{Voelk}}}, \bauthor{\binits{P.L.}
\bsnm{{Biermann}}}, \batitle{{Maximum energy of cosmic-ray particles
accelerated by supernova
  remnant shocks in stellar wind cavities}}.
\bjtitle{\apjl}, \bvolume{333}, \bfpage{L65},
(\byear{1988}). 
\end{barticle}
\endbibitem

\end{thebibliography}


\end{document}